# Ideally Glassy Hydrogen Bonded Networks

J. C. Phillips,

Dept. of Physics and Astronomy, Rutgers University, Piscataway, N. J., 08854-8019

ABSTRACT

The axiomatic theory of ideally glassy networks, which has proved effective in describing phase diagrams and properties of chalcogenide and oxide glasses and their foreign interfaces, is broadened here to include intermolecular interactions in hydrogen-bonded polyalcohols such as glycerol, monosaccharides (glucose), and the optimal bioprotective hydrogen-bonded disaccharide networks formed from trehalose. The methods of Lagrangian mechanics and Maxwellian scaffolds are useful at the molecular level when bonding hierarchies are characterized by constraint counting similar to the chemical methods used by Huckel and Pauling. Whereas Newtonian molecular dynamical methods are useful for simulating large-scale interactions for times of order 10 ps, constraint counting describes network properties on glassy (almost equilibrated) time scales, which may be of cosmological order for oxide glasses, or years for trehalose. The ideally glassy network of trehalose may consist of extensible tandem sandwich arrays.

## I.     Introduction

A few inorganic glasses exhibit extraordinary properties: remarkable mechanical stability, densities of order 90% of their crystalline counterparts, nearly reversible glass transitions, and almost no bonding defects that would create electronic traps. These glasses are composed of stress-free (ideal) networks, and these networks can persist metastably at temperatures below their glass transitions for very long times. In the last 25 years our knowledge of these networks has grown rapidly, both experimentally and theoretically, especially for inorganic chalocogenide and oxide glasses (such as window glass) [1]. At the same time, the idea that H-bonded protein networks exhibit many (almost equilibrated, nearly reversible) properties similar to those of inorganic network glasses has become popular [2-4].

Deductive multiparameter approaches to such ideal glasses, employing standard polynomial Newtonian methods (such as molecular dynamics) encounter severe difficulties even in supercooled liquids (where the viscosity grows exponentially and diverges as $T \rightarrow T_g$), and these difficulties increase in the glass, where relaxation follows the even slower path described by stretched exponentials. In practice these difficulties often limit simulation times to 10 ps. Some simulations of inorganic glasses have circumvented these difficulties by guessing that a good approximation to the glassy network can be obtained by ring disordering an amorphous solid; this guess can be tested semiquantitatively (but with only polynomial, not exponential, accuracy) by comparison with radial distributions observed by diffraction [5], but this method does not predict phase diagrams.

An abstract, parameter-free axiomatic method, based on the variational concepts of Lagrangian mechanics, has provided an excellent guide for experiments on inorganic network glasses, and it was indeed helpful in identifying the intermediate compositional window where non-reversing and ageing effects are small, as well as the internal network stress [1]. This method is hierarchical in nature, and its application involves the general principles of chemical bonding as utilized by Huckel and Pauling. The accuracy of the method in predicting optimized compositions (without using any adjustable parameters) can be as good as 1%. It is summarized in Sec. 2, where examples of Lagrangian constraints show how the method works in the simplest cases of covalent-ionic chalcogenide and oxide network glasses.

The success of the abstract axiomatic method can be tested best by showing that it can be used to identify ideal glasses. For the inorganic network glasses the success has been confirmed in many ways [1], but most of these have not yet been tested for organic glasses. Some organic glasses, such as polymers, are not well described by constraint theory, and to broaden constraint theory to include some cases of hydrogen bonding (Sec. 3) requires careful choices. It turns out that several polyalcohols and saccharides are good candidates for ideal glasses; the reasoning that led us to those choices, as opposed, for example, to polymers, is discussed in Sec. 3, where the hierarchy of H-bonding interactions is inserted into the covalent hierarchical framework. Some macroscopic properties of polyalcohols and saccharides are collected in Sec. 4. It is then straightforward to enumerate (Sec. 5) these interactions for polyalcohols and monosaccharides, but when the procedure is extended to sucrose and trehalose, the results are surprising. Molecular dynamics models in the time range 10 ps − 1 ns partially explain the remarkable biopreservative properties of trehalose, but other factors could be important on a time scale of years. The hierarchical results for trehalose lead to the tandem array network model described in Sec. 6; this model represents a substantial refinement of the two-dimensional percolative model of trehalose films proposed in recent molecular dynamics models of trehalose performed in the time range 10 ps – 1 ns . This model is used to contrast the structure and properties of trehalose with a polyester biopreservative, cutin, in Sec. 6. There we note analogies with extensible tandem arrays of the elastic protein titin and other extracellular matrix and cell adhesion molecules, such as fibronectin, which also contain tandem arrays. It appears that the tandem nature of trehalose networks corresponds well to novel tandem repeats in the cell surface proteins of archaeal and bacterial genomes (Sec. 7). In the Appendix there are some general remarks on the H bonding network of water, especially at interfaces with proteins, polyalcohols, and saccharides,

## 2. Covalent Bond Hierarchies

The simplest case of a covalent glass is a binary one $A_{1-x}B_x$ in which the covalent radii are nearly equal because A and B belong to the same row of the periodic table, for example, A = Ge and B = As or Se. Then the B-A-B and A-B-A bond angles are nearly equal and the bond-bending constraints at the ideal composition are all intact, while there are no constraints on dihedral conformations. This situation is illustrated in Fig. 1(a). The ideal glass-forming condition is ($N_c$ = number of constraints) = $N_d$ = number of degrees of freedom = 3(number of atoms $N_a$), all per formula unit) according to axiomatic mean-field theory [5]. Simulations with space-filling models, based on bond-deleted, ring-disordered space-filling networks, confirmed this condition, and showed a crossover from floppy to stiff networks at the ideal composition, with a density of soft modes nearly linearly decreasing with increasing connectivity [6].

What happens when A and B radii are quite different? This is the case in g-$SiO_2$, where diffraction data show that the width of the O-centered bond angle distribution is much greater than that of the Si-centered distribution. Then the former bond-bending constraints are broken, while the latter bond-bending constraints are intact, as shown in Fig. 1(b). In $(Na_2O)_x(SiO_2)_{1-x}$ alloys with increasing x the Na atoms cut (or form "non-bridging") O-Si bonds, and there is a crossover from broken oxygen bending constraints to intact ones; this crossover actually shows up in the phase diagram as a narrow low-temperature immiscibility gap ($T_c \sim 1000$ K) [7], a transition that was previously unexplained, and which has so far been inaccessible to molecular dynamics simulations (MDS) confined to T > 3000 K. More generally, the effects of space-filling on constraints is subtle and is best determined empirically from traditional structural data (diffraction, Raman, infrared), or even from the phase diagram and the location of the stiffness transition. An important point is that the axiomatic rules are discrete (constraints are nearly always broken or intact in glasses, and are seldom in an intermediate case), and have been refined systematically, in ways that are transferable between situations that are apparently very different, and are conventionally described by completely different (and often quite large) sets of adjustable parameters.

The ways in which phase diagrams and structural data can be interpreted in terms of intact and broken constraints are illustrated in Fig. 2. In underconstrained glasses,

Fig.2(a), there are not enough bond-stretching and bond-bending constraints to exhaust the $3N_a$ degrees of freedom. Thus some of the dihedral angles can be constrained. In the case of g-Se, there are one stretching and one bending constraints per atom, leaving room for one dihedral angular constraint per atom. This makes it possible for g-Se to form long chains (~ 300 atoms) [8], yet remain glassy because of entanglement. As cross-linking Ge or As atoms are added, the number of allowed dihedral constraints decreases, and the chain segments rapidly shorten.

In overconstrained glasses (Fig. 2(b)), the number of bond-stretching and bond-bending constraints is too large, and some redundancies will occur. A simple way for this to happen is for pyramidal or tetrahedral building blocks to share edges, but there are other possibilities, such as replacement of single bonds by double bonds.

Nanoscale phase separation is common in binary glasses. Percolative backbones can consist of molecular units that locally satisfy the ideal glass condition, also describable by isostatic, a term borrowed from hydrodynamics to describe parts of the network that are strain-free. The fraction of the network that is isostatic is variable, and this leads to the formation of a narrow range of compositions with sharp edges that have very favorable properties. The density reaches a plateau, and the glass transition is nearly reversible, and shows little aging [1]. In this reversibility window the network is unstressed, and Raman vibrational frequencies shift linearly with hydrostatic pressure. Outside the window the shifts are small until a threshold pressure is reached, which is interpreted as an internal network pressure [9]. For underconstrained networks, this pressure stiffens soft matrices before it affects isostaic backbones, and for overconstrained networks, it stiffens the isostatic regions before it affects the stiffer overconstrained regions.

## 3. H Bond Hierarchies

Hydrogen bonding D − H …A energies are small (~ 3 kCal/mol) and problematic, partly due to the relative weakness of the interaction. Their electronic components have been studied for a few small molecules [10], while in larger molecules these energies are subject to large dynamical screening corrections, especially those due to the large OH dipoles responsible for the dielectric constant $\varepsilon_0 \sim 80$ of water.  Thus recent models of H bond interactions in proteins are empirical and are based on complex statistical analysis designed to differentiate H bonding interactions with both peptide backbones and amino acid side groups [11]. However, one may still expect the usual hierarchy of interaction energies, with E(O − H…A) > E(N − H…A) > E(C − H…A), and E (bond stretching) > E (bond bending).   For the polyalcohols considered below, where the complexities produced by N lone pairs are avoided, this leads to the simple hierarchy shown in Fig. 3(b).   It is far from obvious that the C − H…A bending energies are larger than the covalent dihedral energies; these two could be grouped together without changing some of our results.

Polymers may avoid crystallization because of entanglement, and similarly molecular glasses may form because of steric hindrance.  The polar nature of H bonds suggests that dynamical interactions with local electrical fields to form H bond networks can compete with steric hindrance in promoting the glass-forming tendency, and be more easily quantified. Thermal expansion at constant pressure ($\alpha_P = V^{-1}(\partial V/\partial T)_P$) and at constant dielectric relaxation time ($\alpha_\tau = V^{-1}(\partial V/\partial T)_\tau$) near the glass transition temperature ($\tau = 1$ s) provide an easy way [12] to gauge quantitatively whether or not H bonding is critically enhancing the glass-forming tendency. (The results are quantitatively similar for $\tau = 1000$ s, or when $\tau$ is replaced by the viscosity $\eta$ [13].) The measured values of the ratio $\alpha_\tau/\alpha_P$ are near unity (they range from 0.6 to 2.8) for 15 molecular and polymer glass-formers, but are 6 and 17 for the strongly hydrogen bonded polyalcohols sorbitol and glycerol, respectively. H bonding is thus less sensitive to pressure, and more sensitive to temperature, than covalent bonding, and is more dominant, and therefore probably more easily quantified, in polyalcohols than in polymers.

## 4. Polyalcohols and Saccharides

Many polyalcohols are good glass formers, so they are a good starting point for broadening constraint theory to include hydrogen bonding. The glass transition temperatures $T_g$ of a few polyalcohols and sugars are listed in Table I, together with the extrapolated slopes of reduced viscosity (fragilities m) on reduced temperature scale, that is, m = dlog $\tau$/dlog($T_g$/T)$_{T = Tg}$ [14]. Saccharide data [15, 16] are also included, as these also have strong H-bonding interactions.

Clustering is a common property of good glass formers; a quantitative measure of clustering is the ratio of the minimum in Raman scattering intensity before the Boson peak to the value at the peak, denoted by $R_1$. In the well-known molecular glass former, sterically hindered orthoterphenyl (OTP), this peak is absent, so in Table I the value $R_1$ = 1.0 (OTP). Given the excellent glass-forming tendency in strongly H-bonded glycerol, one would expect to see a strong Boson peak, as shown in Table I. Even stronger Boson peaks (smaller $R_1$) are seen in the bioprotective glass formers sucrose and trehalose [17].

The stretched exponential relaxation function $\exp[-(t/\tau)^\beta]$ provides another measure of the many-particle interactions in glasses, as perturbed by a variety of probes, through the dimensionless fraction $\beta$ [18]: smaller values of $\beta$ correspond to longer range interactions, as short-range interactions alone give $\beta = 3/5$, while a mixture of short-and long-range interactions gives $\beta = 3/7$ [19]. The most accurate values of $\beta$ (to a few per cent) are usually obtained in the time domain, and the ones quoted in Table I are largely from such measurements [18,20]. Similar values (with larger uncertainties) can be obtained from dielectric relaxation or scanning calorimetry [15].

The many-particle interactions affect m and $\beta$ independently: in sorbitol-glycerol mixtures, m varies smoothly and monotonically, but $\beta$ does not [21]. From the values of $\beta$ we see that glycerol relaxes via short-range interactions, while sorbitol is affected by a mixture of short-and long-range interactions, presumably due to formation of chain bundles. A rough linear correlation between m and $\beta$ has been suggested [22], and both sorbitol and glycerol lie in the "allowed" band, but near equal weight percents the mixture lies far outside this band, with $\beta$ "anomalously" small (exceptionally wide distribution of relaxation times). This suggests that there is a dynamically driven short-range ordering of nearly spherical glycerol with sorbitol chain segments in those mixtures (possibly with glycerol inserted between the sorbitol chains), which opens a new range of anisotropic relaxation channels unavailable in the pure materials. Thus while large values of the fragility m often describe a glass which is strongly associated just above $T_g$, there are several other possibilities.

### 5. Constraint Counting in Polyalcohols and Saccharides

There have been several studies of glycerol by molecular dynamics simulations (MDS) [23,24]; the molecular structure is shown in Fig. 4. In the crystal [23] each molecule is bound to four neighbors by six hydrogen $D - H \ldots A$ bonds, three with $D = O$ and $A = C$, and three with D and A reversed. The bond lengths are near normal values, and the $D - H \ldots A$ bond angles are fixed by minimization of the overall electrostatic and torsional energies in the context of the crystalline space group. Thus, as expected, one cannot learn much about the effective H-bonding stretching and bending constraints from the crystal structure, but the MDS results [23] for the glass transition are much more interesting. The predicted value of $T_g$ is in good agreement with experiment, while the width $\Delta T_g$ of the glass transition is greatly overestimated (by about a factor of 10 for the faster cooling rate of 200 K/ns, and by about a factor of 5 for the slower cooling rate of 100 K/ns), indicating that many weaker H-bonding constraints are intact in the glass that are not attained in the MDS on these time scales.

Constraint counting for glycerol gives the following results: ($N_A = 14$); Constraints **[running total]**: intramolecular stretching constraints: 13 **[13]**. intramolecular bending $(2N - 3$, where N is the number of single bonds): 3C, N = 4, $3(2N - 3) = 15$ **[28]**, 3O, N = 2, $3(C\text{-}O\text{-}H) = 3$ **[31]**. Intermolecular: stretching H…, 8 **[39]**; bending, $3(O\text{-}H…)$ **[42]**. (The C-H… bond has a stretching constraint, but it is too weak to have a bending constraint. Stated differently, the intact-broken gap in the constraint hierarchy at $T = T_g$ lies between the O-H…(intact) and C-H…(broken) bending constraints. Total number of intact constraints: **[42]** $= 3 N_A$. Thus glycerol is an ideal glass, the prototypical "viscous solvent". Presumably the relatively weak O-H… bending constraints are not optimized by MDS, which is why MDS overestimates $\Delta T_g$. Dielectric relaxation and NMR also lead to the conclusion that that the O-H…constraints are intact in glycerol [14,25], while the C-H… bending constraints are broken. There is no evidence in glycerol for a large "excess wing" in the dielectric relaxation of glycerol. This is consistent with the "forced relaxation" model of dielectric relaxation in ideal glasses [19].

Constraint counting for sorbitol, on the other hand, Table II, shows that there are not enough O-H… constraints to freeze the glass, but that if all the C-H… bending constraints were intact, the glass would be overconstrained. Dielectric relaxation and NMR data [14,25] for sorbitol show that the O-H… and C-H… bending constraints behave similarly; in other words, because the ideal glass-forming condition is not satisfied, at the molecular level sorbitol is more nearly amorphous than glassy. (Even in glycerol, where there is a difference between the O-D… and C-D… spin lattice relaxation times, the ratio is only a factor of 2. This can be compared to the factor of 9 difference between the corresponding dipole moments predicted by a standard molecular dynamics program (AMBER) [23]. One model for the relaxation would utilize fluctuations of the polarization of the neighborhoods of the dipoles to relax them; if this polarization depended only on the dipole in question, then the rate would scale with that dipole's moment. In fact, the polarization depends on many other local dipole moments as well, which is why the rates differ only by a factor of 2.) This means that the macroscopic glass transition in sorbitol involves entanglement of a polymeric nature; the rapid temperature dependence of such entanglement as $T \rightarrow T_g$ could explain why sorbitol is the most fragile glass listed in Table I. Thus sorbitol is not an ideal glass, and this explains why its dielectric relaxation spectrum shows a large "excess wing" [14,25]. Similarly, crossovers in the high-frequency glassy behavior of oligo(propylene glycol) dimethyl ethers with variable chain length [26] apparently arise from polymeric conformations suggestive of collective entanglement in the H-bond network not describable by microscopic H-bonding hierarchies.

Turning now to the monosaccharides (Table II), we see that glucose and fructose merely have different conformations, and that neither is an ideal glass former at the microscopic level. Trehalose is a fully symmetrical disaccharide (see Fig. 5) with unique properties: it has been found in large quantities in organisms (algae, bacteria, fungi, insects, invertebrates, and yeasts as well as a few flowering plants) that are able to survive extreme external stresses such as high or very low temperatures or periods of complete drought up to 120 years (anhydrobiosis). These qualities led to the suggestion [27], now apparently prescient, that trehalose forms a glassy structure around embedded biomolecules and inhibits thus the denaturization due to formation of ice crystals. Table I shows that trehalose has the highest $T_g$, the best-developed Boson peak, and the widest distribution of relaxation times of alcohols and saccharides, so we expect it to show ideal glass forming tendencies at the microscopic level.

At first sight, the constraint count for trehalose seems to show (Table II) that it cannot be an ideal glass, as the number of constraints including only C-H… stretching is too small, whereas when the C-H… bending constraints are added, the entire molecule is overconstrained. However, a closer look shows that if only half of the C-H… bending constraints are intact, the ideal glass condition is indeed satisfied. This would be the case if the C-H… bending constraints are intact for one flap (say A) and broken for the other (B). This very simple observation has far-reaching consequences, as we shall see below.

Like glycerol, but unlike sorbitol, dielectric relaxation and ultrasonic velocity dispersions in trehalose and maltose are featureless [28], as one would expect for ideal glasses. There are small differences between trehalose and maltose, with the activation energy and hydration number being about 20% and 10% larger, respectively, for trehalose than for maltose [28]. NMR data [29] show that the rotational barrier for the glycosidic torsional angles ($\varphi$ and $\varphi'$ in Fig. 5) is $\sim 700$ K when it is calculated relative to gas-phase interactions; this value must actually be less than $T_g$, so the glassy interactions must be smaller than those in the gas phase, because of dynamical screening effects. Another interesting feature of the NMR data is that the broadening of the lines produced by the distributed torsional angles is large both for the H's attached to the C's adjacent to the bridging O and involved directly in the torsional motion, and the H's in the ethanol ($H_2COH$) side group ($C_{11,12}$ in Fig. 5).

We conclude with analyzing sucrose, which is similar to trehalose, but with H at the C2 position replaced by $CH_2OH$ [30]. This does not change the constraint count much, but the symmetry of the two flaps is destroyed, and the ideal glass condition is no longer satisfied. An intermediate case is maltose, which combines rings with the glucose and fructose conformations, instead of two glucose conformations, thus leaving the constraint count ideal at the trehalose value, but breaking the symmetry. [30] suggested that the difference in the aqueous bioprotective properties of trehalose, maltose, and sucrose can be explained by two-dimensional percolative models. In the context of the unique bioprotective functionality of trehalose, such models lack specificity. These models are refined by the extended intermolecular trehalose structural model discussed next.

## 6. Tandem Bilayer Trehalose Films

Because the ideal glass condition is satisfied for trehalose when half of the C-H… bending constraints are intact, one is led naturally to the staggered tandem bilayer model shown schematically in cross section in Fig. 6(a). Rigid pyranose rings are alternately more and less tightly paired. Such a binary alternation has many favorable bioprotective features: for instance, it combines stability with extensibility, as we shall see. Fig. 6(b) shows in cross section how the ethanol side groups are part of the alternating structure; a similar alternation occurs in and out of the plane. This is possible because of the mirror symmetry of the two glucose conformations. In maltose, where glucose and fructose conformations are combined, both of the ethanol groups are either inside or outside, and this alternation is not possible. This partially explains the superior bioprotective properties of trehalose. The alternation of the ethanol groups from inside to outside in trehalose is also consistent with the observed ethanol broadening and shifts seen by NMR [29].

One can now ask which factor is decisive in stabilizing the tandem sandwich trehalose structure: the partitioning of the C-H… bending constraints, or the alternation of the ethanol side groups? With respect to the structure itself, this question cannot be decided, as both are present. However, there is by now considerable evidence that the special properties of trehalose persist in aqueous solution [30]. As shown in the Appendix, water interfaces well with planar hydrophilic substrates, forming a collective, stress-free monolayer; this agrees well with the partitioning of the C-H… bending constraints. By contrast, the ethanol water interactions are no longer planar, and the symmetry of the interface is destroyed near the ethanol units, which are also local units that do not favor collective interactions across an entire glucose flap.

The favorable mechanical properties of binary alternating tandem structures are best illustrated not by calculation, but by direct analogy with evolutionarily designed networks. Thus the much-studied giant skeletal muscle protein titin comprises a tandem array of fibronectin type III and immunoglobulin domains, which are structurally similar 7-strand beta-sandwiches [31]. Many extracellular matrix and cell adhesion molecules, such as fibronectin, contain tandem arrays of fibronectin type III domains. It was suggested [31] that both single molecules and matrix fibers should have elastic properties similar to titin. The present one-dimensional tandem model for trehalose can itself be extended to form a two dimensional staggered checkerboard film (Fig. 6(c)), with essentially complete bioprotective properties. (Of course, this structure, which appears to be crystalline, can be strongly disordered in the amorphous phase, without sacrificing most of its favorable packing features.)

Cutin forms interesting bioprotective films that are less glassy than trehalose. Cutin is a support biopolyester involved in waterproofing the leaves and fruits of higher plants, rendering them shiny, regulating the flow of nutrients among various plant cells and organs, and minimizing the deleterious impact of pathogens [32]. Infrared and Raman data show that cutin consists of fatty polymeric acids with few or no stabilizing rings. It can be destabilized by glycerol [33].

## 7. Protein-Trehalose Interfaces

One more step is necessary to demonstrate the specificity of the tandem bylayer model shown in Fig. 6: the substrates themselves should exhibit tandem patterns. In fact, computer searches of genomes of the cell surface proteins of archaea and bacteria have identified many tandem repeats, including one in a single-stranded DNA-binding protein domain that was presumably present in the common ancestor to all three major branches of life, archaea, eukarya, and bacteria [34], with at least four tandem repeats (~ 100 amino acids). Several longer repeats (up to seven) have recently been found [35]. The adaptability of tandem repeats supports cooperative folding in multi-domain proteins [36] and favors evolutionary "agility" [37, 38]. The folding of sandwich-like proteins involving interlocked pairs of neighboring β strands exhibits bimodal behavior similar to trehalose [39]: half of the residues form native-like residues in the folding transition state, whereas the other half are absent from the folding state, but present in the native state. It appears possible that the tandem structure of trehalose (Fig. 6) could be the origin of the tandem repeats that appear to dominate much early evolution, and still may play an important part today in forming paths by which regulatory sequence can change, yet preserve function. The reader may note that the tandem model for trehalose has been derived here without the use of adjustable parameters, or elaborate statistical fits of undetermined reliability.

I am grateful To Profs. M. Descamps and L. Cordone for discussions on trehalose.

## Appendix: Water and Its Interfaces

Realistic biological interfaces are mediated by at least a monolayer of water. Although the freezing point of water is low, and its glass temperature is presumably much lower than that of glycerol, theory should be concerned with the structure of water in confined geometries. MDS shows [39] abnormal behavior of a few layers of water between hydrophobic solutes, but most protein surfaces are hydrophilic, and trehalose is strongly hydrophilic. Porous Si is weakly hydrophilic, and in this confined geometry $T_g$ already increases to 220 K [41]. In fact, water probably does form networks when confined as a few monolayers between strongly hydrophilic surfaces. However, constraint theory says that these networks are *stress-free* [1], which means that trehalose films can indeed carry out their bioprotective function.

The ideality of protein – water interfaces is implicitly guaranteed by evolution, but the nature and origin of this ideality remains one of the central explicit problems in biology. Recently emphasis has shifted from continuum models of solvation of small molecules [42] to the glassy network nature of the first hydration shell around proteins [2,43]. X-ray and neutron scattering in $H_2O$ and $D_2O$ solutions showed that the first hydration shell around proteins has an average density ~ 10% larger than that of the bulk solvent [43,44]. MDS has shown that the strongest coupling is between the hydration shell and the bulk water, while the coupling of the hydration shell to the protein substrate is weak [45], a conclusion supported by experiments [2].

Constraint theory provides an independent test of the ideality of the first hydration shell. With maximal hydrogen bonding, O is effectively 4-fold coordinated, and H is effectively 2-fold coordinated. Thus $H_2O$ at low T becomes topologically isomorphic to $SiO_2$. This explains the similarities of the (P,T) phase diagram of ice at high P to that of $SiO_2$. Bulk $SiO_2$ is an ideal glass [7], and the $Si/SiO_2$ interface is the most perfect substrate/glass interface known (defect concentration $< 10^{-4}$). An Si*O* monolayer is sandwiched between an $SiO_2$ overlayer and the Si substrate, and one can now calculate the constraints by a symmetry argument [46]. The Si* atom in the SiO monolayer forms on the average two bonds with the Si substrate, and two bonds with the $SiO_2$ overlayer. This gives Si-Si*-Si and O-Si*-O bending constraints, and the angle between the normals to the intact Si-Si*-Si and O-Si*-O planes supplies one further bending constraint. There are 4/2 stretching constraints for Si*, and 2/2 stretching constraints for O*, giving a total of 6 = $2N_A$ bending and stretching constraints for the Si*O* monolayer, which forms a perfectly glassy interface. Repeating this argument for the hydrogen bonded first hydration shell O*H* between a protein substrate and an $H_2O$ overlayer, we conclude that the hydrogen bonded first hydration shell O*H* is also an ideally glassy (fully off-lattice, stress-free [1]) interfacial layer. The 10% density enhancement can be regarded as a secondary effect associated with the relative dominance of hydrophilic interactions at the surface of the protein compared to hydrophobic ones in its core, where some water is still buried.

| Material | Formula | $T_g$ (K) | m | $R_1$ | $\beta_t$ | $\beta_{fd}$ |
|---|---|---|---|---|---|---|
| glycerol | $HCOH(H_2COH)_2$ | 187(2) | 57 | 0.60 | 0.60 | 0.65 |
| sorbitol | $(HCOH)_4(H_2COH)_2$ | 268 | 128 | - | - | 0.37 |
| glucose | $C_6O_6H_{12}$ | 308 | 70 | - | - | 0.64 |
| fructose | $C_6O_6H_{12}$ | 274 | 48 | - | - | 0.50 |
| trehalose | $(C_6O_5H_{11})_2O$ | 396 | 107 | 0.34 | 0.38 | 0.30 |
| sucrose | $(C_6O_5H_{11})\ (C_6O_5H_{10}CH_2OH)O$ | 233 | 60 | 0.48 | - | - |
| orthoterphenyl | $C_6H_4(C_6H_5)_2$ | 243 | 81 | 1.0 | 0.62 | 0.51 |

Table I.  Glass-forming tendencies and formulae of polyalcohols and saccharides.  Note that $T_g$ of glucose (sucrose, trehalose) is close to (well below, well above) the operating temperature of proteins.  Here $\beta_t$ and $\beta_{fd}$ refer to stretching fractions measured in the time (more accurate, $\pm 0.02$) and frequency (dielectric relaxation, less accurate $\pm 0.15$, because of narrow band convolution uncertainties [19]) domains.  In any case, the observed values of $\beta$ suggest short-range forces in glycerol, and a mixture of short- and long-range forces in trehalose.

| Material | $3N_A$ | $N_c$ | $(N_c + 2N_{OH})$ | $(N_c + 2N_{OH} + N_{CH})$ | | $(N_c + 2N_{OH} + 2N_{CH})$ |
|---|---|---|---|---|---|---|
| glycerol | **42** | 31 | 37 | **42** | | 47 |
| sorbitol | **78** | 55 | 67 | 75 | | 83 |
| glucose | **72** | 50 | 60 | 67 | | 74 |
| fructose | **72** | 50 | 60 | 67 | | 74 |
| trehalose | **135** | 98 | 114 | 128 | **(135)** | 142 |
| sucrose | **147** | 106 | 124 | 139 | | 154 |

Table II. Constraint counting in polyalcohols and saccharides. The number of degrees of freedom per molecule is $3N_A$, where $N_A$ is the number of atoms in the molecule (frozen in the ideal glass, hence ideally no translational or rotational degrees of freedom for each molecule), the number of covalent bond-stretching and bond-stretching constraints is $N_c$, and the last three columns consider three possibilities: only OH bond stretching and bending, adding CH stretching, and adding CH stretching and bending. Note that quite different results would be obtained from a mechanical spring model with 6 (rather than 5) bending constraints/tetrahedral C; in the case of glucose, $N_c + 2N_{OH} + N_{CH}$ would be increased to $76 > 3N_A = 72$, making glucose and all sugars strongly overconstrained rather than marginally constrained, with glycerol ($N_c + 2N_{OH} + N_{CH} = 45 > 3N_A = 42$) and trehalose ($N_c + 2N_{OH} + N_{CH} = 147 > 3N_A = 135$) no longer being ideal glasses. This could be one of the reasons that the simple results listed in this Table have not been obtained previously. Another even simpler possibility is that Newtonian forces are usually displayed in vibrational models, rather than Lagrangian interactions. For two-body stretching forces, the counting is the same for both pictures, but when three body bending bending forces are added, the Newtonian count is misleadingly large, making it difficult to recognize the structural origins of floppy modes.

**Figure Captions.**

Fig.1.  Typical constraint hierarchies for inorganic network glasses.  (a)  When $N_c <$ $3N_a$, all of the marginal bond-bending constraints will be satisfied, and some of the dihedral degrees of freedom will be floppy.  (b) When $N_c > 3N_a$, some of the marginal bond-bending constraints will not be satisfied, or there will be redundancies, for instance, edge-sharing tetrahedra.

Fig. 2.  Size dependence of constraint hierarchies for inorganic network glasses.  (a) In chalcogenide and some silicate glasses, the average sizes of cations and anions are approximately equal.  Then because the networks are charge-ordered, the condition for ideal glasses is satisfied at or near compositions where all bond-bending constraints are satisfied, and all dihedral angles are widely distributed (floppy).  This is the simplest case.  (b)  When the cation and anion sizes are very different, as in $SiO_2$, the bond-bending constraints can be intact around the larger ion, and broken around the smaller one.

Fig. 3. Comparison of covalent network glasses and polyalcohol glasses.  (a) the ideal covalent case,repeated from Fig. 2(a).   (b)  Polyalcohols contain covalent cores and $D - H \dots A$ bonds, where D and A can be C or O.  In the absence of specific factors (such as small or large sidegroups) the most likely constraint ordering is the one shown here.

Fig. 4.  The molecular structure of glycerol [24].

Fig. 5.  The molecular structure of trehalose.  The H atoms (not shown) complete the covalent intramolecular  network, so that each carbon (oxygen) atom is four- (two-)

fold coordinated. Intermolecular hydrogen bonds can attach to all the carbon and oxygen atoms, except for the bridging oxygen O1 [24]. Sucrose differs from trehalose by replacement of the H (not shown) at C2 by $CH_2OH$.

Fig. 6. (a) Cross section of tandem sandwich structure proposed for trehalose. The double lines indicate a strongly bound pair of glucose rings, which alternate with weakly bound pairs (single lines). (b) Enlarged view of structure. The ethanol side groups lie between (outside) the rings for weakly (strongly) bound pairs. (c) Successive sandwiches can be staggered for better packing.

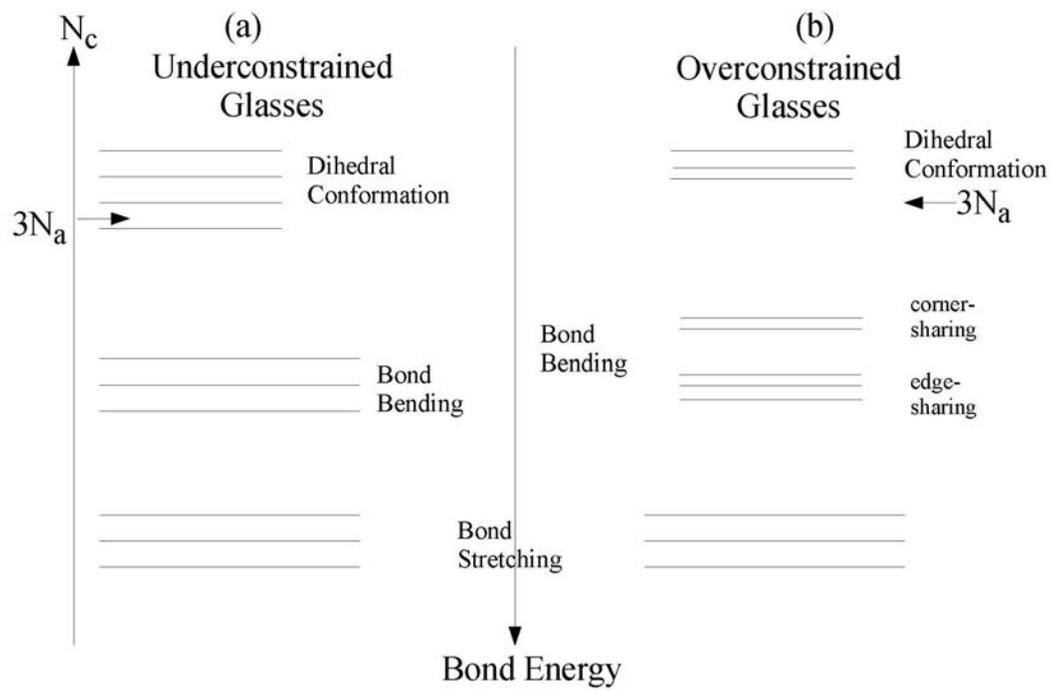

Fig.1

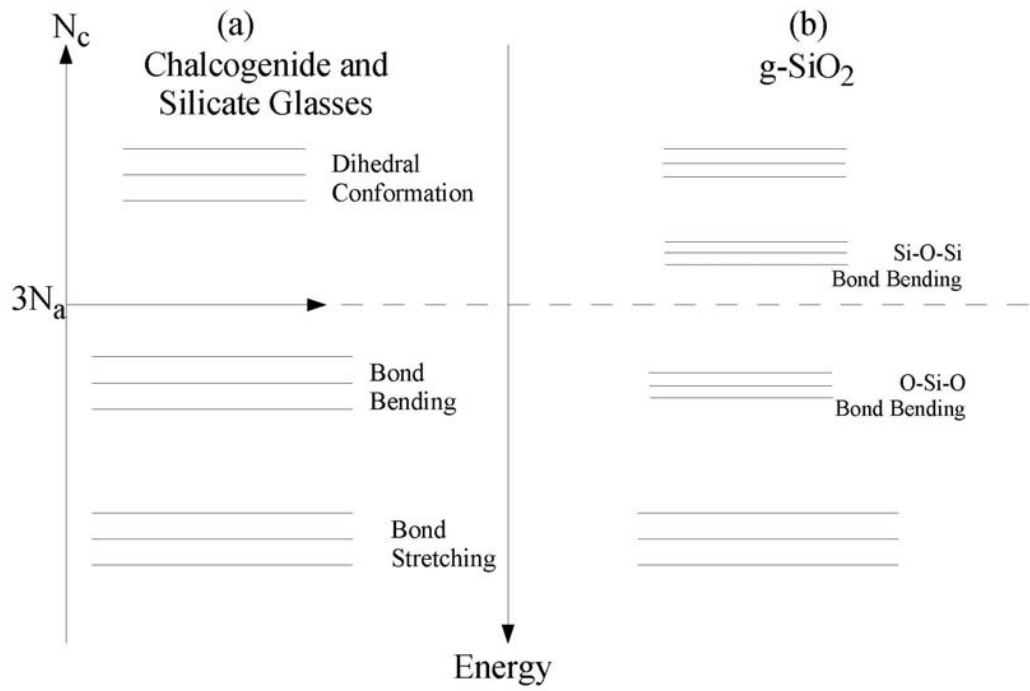

Fig. 2

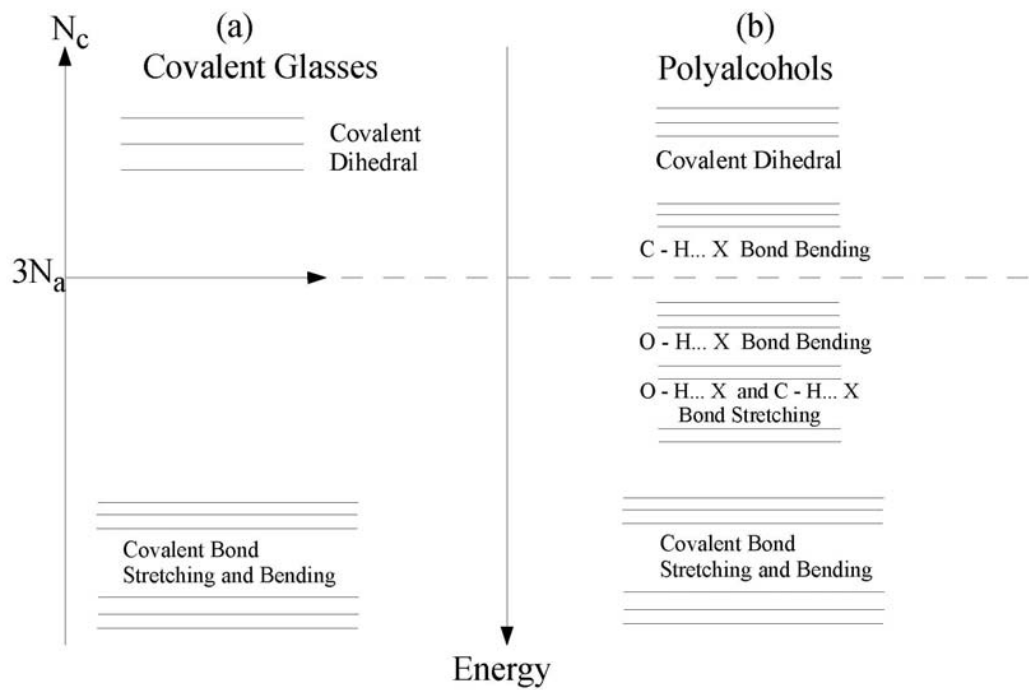

Fig. 3.

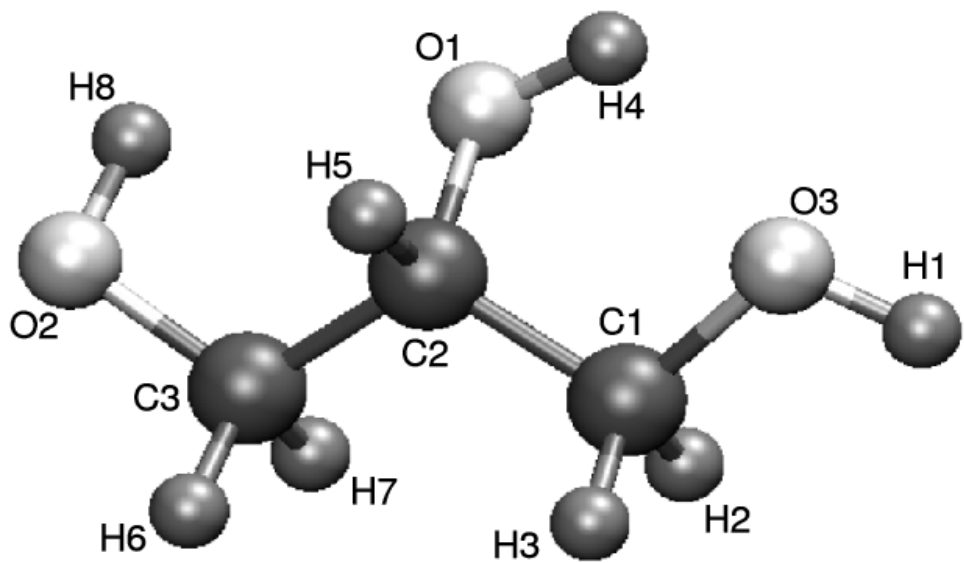

Fig. 4.

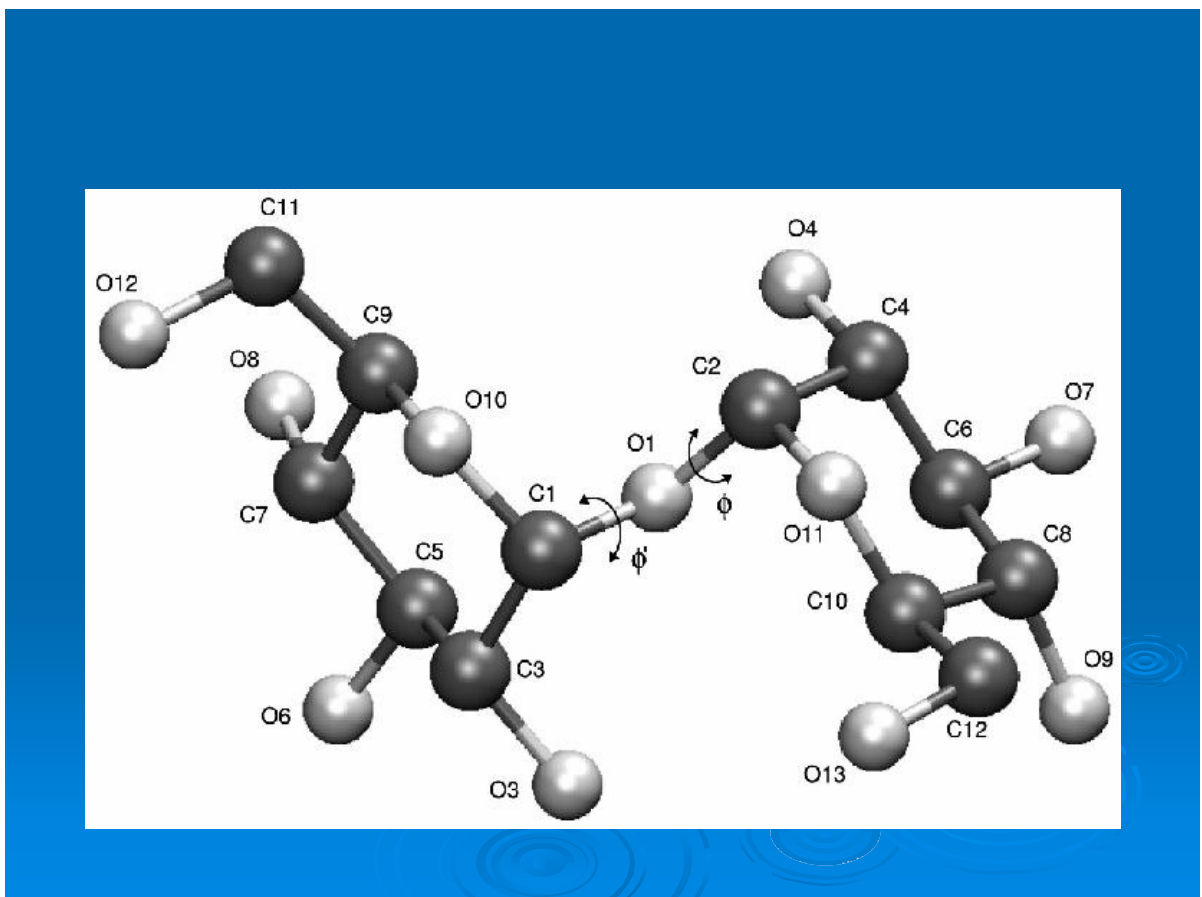

Fig. 5.

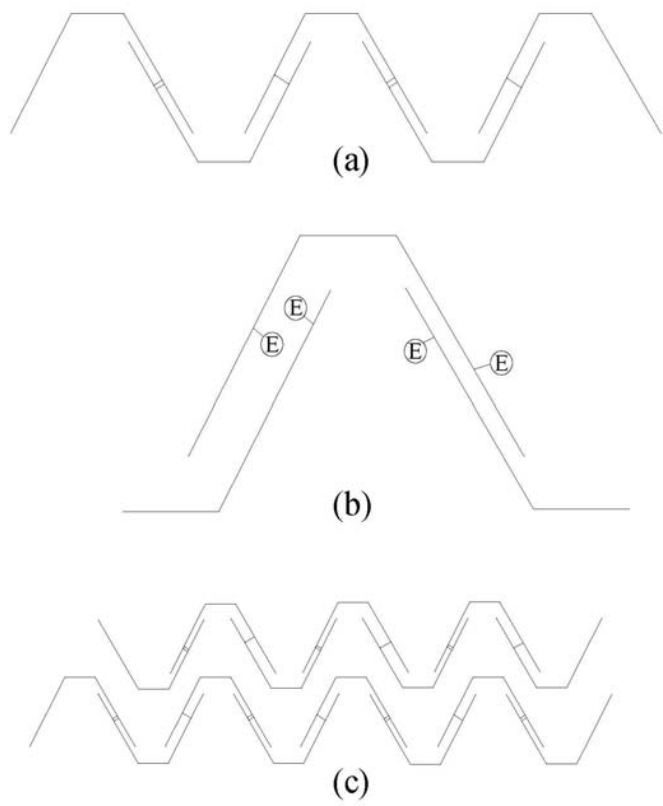

(a)

(b)

(c)

Fig. 6.